# Patents as Instruments for Exploring Innovation Dynamics:

# Geographic and Technological Perspectives on "Photovoltaic Cells"

*Scientometrics* (**in press**)

Loet Leydesdorff,[a]* Floortje Alkemade,[b] Gaston Heimeriks,[c] and Rinke Hoekstra[d,e]

**Abstract**

The recently developed Cooperative Patent Classifications (CPC) of the U.S. Patent and Trade Office (USPTO) and the European Patent Office (EPO) provide new options for an informed delineation of samples in both USPTO data and the Worldwide Patent Statistical Database (PatStat) of EPO. Among the "technologies for the mitigation of climate change" (class Y02), we zoom in on nine *material* technologies for photovoltaic cells; and focus on one of them ($CuInSe_2$) as a lead case. Two recently developed techniques for making patent maps with interactive overlays—geographical ones using Google Maps and maps based on citation relations among International Patent Classifications (IPC)—are elaborated into dynamic versions that allow for online animations and comparisons by using split screens. Various forms of animation are discussed. The longitudinal development of Rao-Stirling diversity in the IPC-based maps provided us with a heuristics for studying technological diversity in terms of generations of the technology. The longitudinal patterns are clear in USPTO data more than in PatStat data because PatStat aggregates patent information from countries in different stages of technological development, whereas one can expect USPTO patents to be competitive at the technological edge.

[a] *corresponding author; Amsterdam School of Communication Research (ASCoR), University of Amsterdam, Kloveniersburgwal 48, NL-1012 CX Amsterdam, The Netherlands; tel.: +31-20-6930565; fax: +31-842239111; email: loet@leydesdorff.net.
[b] Department of Innovation Studies, Faculty of Geosciences, Utrecht University, Heidelberglaan 2, NL-3584 CS Utrecht, The Netherlands; tel.: +31-30-253 5410; email: F.Alkemade@uu.nl
[c] Department of Innovation Studies, Faculty of Geosciences, Utrecht University, Heidelberglaan 2, NL-3584 CS Utrecht, The Netherlands; tel.: +31-30-253 7802; email: gheimeriks@gmail.com .
[d] Department of Computer Science, VU University, De Boelelaan 1081, NL-1081 HV Amsterdam, The Netherlands; tel.: +31-20-598 7752; email: rinke.hoekstra@vu.nl ;
[e] Faculty of Law, University of Amsterdam, Amsterdam, The Netherlands.

## 1. Introduction

Patents are framed in different contexts: in addition to being among the outputs of the production system of knowledge, patents can also serve as input to the economic process of innovation. Furthermore, intellectual property in patents is legally regulated, for example, in national patent offices (e.g., Granstrand, 1999). Patents reflect these different contexts in terms of attributes: names and addresses of inventors and assignees provide information about the locations of inventions; patent classifications and claims within the patents can be used to map technological developments; citations provide measures of impact and value, etc. (e.g., Hall *et al*., 2002; Porter & Cunningham, 2005). Can patent analysis and patent maps provide us with an analytical lens for studying the complex dynamics of technological innovations? (e.g., Jaffe & Trajtenberg, 2002; Balconi *et al*., 2004; Feldman & Audretsch, 1999; Mowery *et al*., 2001).

In this study, we argue that a further development of methodologies is required more than of theories when one understands technologies as complex adaptive systems. The various contexts provide different selection environments that are further explored with the development of the technology. The diffusion of a new technology in different dimensions may vary in terms of the rate and the directions.

In the case of small interference RNA (siRNA), for example, we found in a previous study (Leydesdorff & Rafols, 2011) that the initial discovery was academic and published in *Nature* (Fire *et al*., 1998). After a few years, however, the centers of preferential attachment shifted from the academic inventors to institutional centers of excellence in metropolitan areas such as London, Boston, and Seoul. A spin-off company (Alnylam) was created by MIT and the Max

Planck Society (in 2002) in order to secure the revenues of a number of patents. However, economic exploitation of the technology as a reagent became more attractive than as a diagnostic tool when the transition from *in vitro* to *in vivo* encountered problems (Lundin, 2011). Accordingly, the center of patenting shifted to Denver, Colorado during the 2000s (Leydesdorff & Bornmann, 2012). In the meantime, the academic research front shifted focus from "small interference RNA" to "micro interference RNA" (Rotolo *et al*., in preparation).

The example illustrates that in order to appreciate the complexity of innovation processes and understand the emerging and evolving patterns, one needs instruments to study the different dimensions and the interactions among them over time and in relation to one another. In this study, we build on the recent development of geographical maps of patents and maps in terms of patent classes as different projections (Leydesdorff and Bornmann, 2012; Leydesdorff, Kushnir, and Rafols, 2012). We extend the static maps with a methodology to study the evolution of inventions over time in the different dimensions. For example, using the proposed methodology one can overlay the networks of co-inventors on a Google Map or analyze these networks using measures from social network analysis (Breschi and Lissoni, 2004). Different dimensions and dynamics can thus be distinguished and then related. Can the co-evolution in different dynamics be grasped in order to show two or more dynamics in parallel by using split screens?

Several teams have generated patent maps and overlays for patent classes (Kay *et al*., in press; Schoen *et al*., 2012). However, our main objective is to make these overlays *interactive* so that one can use them as versatile instruments across samples gathered for different reasons. In our opinion, one must be able to change the focus in order to capture the resulting dynamics. In summary, we add to the previous mappings and overlays: (*i*) the dynamics by using time series, (*ii*) the social networks, and (*iii*) options to consider more than a single dynamics concurrently—but not necessarily synchronously—using split screens (Leydesdorff & Ahrweiler, in press).

As a case, we focus on a specific material technology for photovoltaic cells ($CuInSe_2$), but our aim is to demonstrate the methodology and further develop the overlay techniques for sequential years into animations. Accompanying websites provide instructions for using the instruments for other sets.[1]

**2. Patent data**

Despite the well-known limitations (e.g., Archibugi & Pianta, 1996; OECD, 2009), patents can be used for analyzing patterns of invention along the dimensions of locations, technology classes, and organizations. The freely accessible interface of the United States Patent and Trade Office (USPTO) allows us to download sets of patents in batch jobs on the basis of composed search strings, and additionally to track their citation rates. An SQL-script was furthermore developed that enables the user to draw patents similarly from the Worldwide Patent Statistical Database (PatStat) of the European Patent Office (EPO).

The PatStat database includes patents of more than 80 patent offices worldwide (including USPTO, EPO, and the Japanese Patent Office), but access to this database requires institutional subscription. The expectation is that PatStat, because of its broad coverage in terms of patent offices, can inform us about networks at national or regional levels that may be coupled to developments in USPTO to varying extents. The US market provides a highly competitive environment, whereas technologies can also be further developed in niche markets. The latter may be more visible in PatStat data than USPTO data.

---

[1] For using USPTO patents, see at http://www.leydesdorff.net/software/patentmaps/dynamic ; and for PatStat data analogously at http://www.leydesdorff.net/software/patstat .

## 3. Cooperative Patent Classifications of "Photovoltaic Cells"

On the 1st of January 2013, USPTO and EPO introduced a new system of Cooperative Patent Classifications (CPC)[2] that unlike existing patent classifications (such as International Patent Classifications IPC, and its American or European equivalents), can also be indexed with a focus on emerging technologies using specific tags in the new Y-class (Scheu *et al*., 2006; Veefkind *et al*., 2012). Whereas the previous classification systems have grown historically with the institutions, and combine patents that cover product and process innovations at different scales, the classification in terms of CPC adds technological classes from the perspective of hindsight under the category "Y".

EPO first experimented with the class Y01 as an additional tag for nanotechnology patents (Scheu *et al*., 2006), while USPTO tried to accommodate nanotechnology into a subclass 977 of its existing classification system. "Y01" was subsequently integrated into IPC v8 as class B82. More recently, a new CPC tag for emerging technologies was developed as Y02: "Climate Change Mitigating Technologies." In the meantime, these new classifications have been backtracked into the existing databases for indexing.[3] The tag and its subclasses are now operational in both USPTO[4] and PatStat data.[5]

---

[2] See for more information about CPC at http://www.cooperativepatentclassification.org/index.html .

More than 150,000 patents are tagged with Y02 in USPTO, among which 5,021 US patents with the search string cpc/y02e10/54$ for *material technologies* in photovoltaic (PV) cells (cf. Peters *et al*., 2012; Shibata *et al*., 2010). In terms of CPC, these technologies are further subdivided into nine specific technologies as shown in Table 1.[6]

| CPC | Description | USPTO[7] | PATSTAT[8] |
|---|---|---|---|
| **Y02E 10/541** | **CuInSe2 material PV cells** | **422[9]** | **3428** |
| Y02E 10/542 | Dye sensitized solar cells | 532 | 8903 |
| Y02E 10/543 | Solar cells from Group II-VI materials | 294 | 2396 |
| Y02E 10/544 | Solar cells from Group III-V materials | 850 | 4116 |
| Y02E 10/545 | Microcrystalline silicon PV cells | 146 | 1071 |
| Y02E 10/546 | Polycrystalline silicon PV cells | 262 | 1709 |
| Y02E 10/547 | Monocrystalline silicon PV cells | 1158 | *n.a.* |
| Y02E 10/548 | Amorphous silicon PV cells | 742 | 5374 |
| Y02E 10/549 | organic PV cells | 1340 | *n.a.* |

**Table 1**: Nine material technologies for photovoltaic cells distinguished in the Cooperative Patent Classifications (CPC).

We focus in this study on developing the relevant instruments using the first subclass Y02E10/541 that covers "CuInSe$_2$ material PV Cells." CuInSe$_2$ is used in thin-film solar cells; thin-film solar cells are an emerging technology and are expected to be a dominant photovoltaic (PV) technology in the future (Unold and Kaufmans, 2012). Although this technology has only a small share of the market, it continues to attract most of the funding for R&D among the material technologies for photovoltaic cells (*ibid.*, p. 12).

---

[3] At the date of this research (August-October 2013), this backtracking had been completed for USPTO data, but not for the then current version of PatStat (April 2013), for two of the nine classes here under study (see Table 1 below). The USPTO envisages replacing the US Patent Classification System (USPC) with CPC during a period of transition to 2015; at EPO, however, the European classification ECLA has already been replaced with CPC.

[4] The Y02-class follows up on the "Pilot Program for Green Technologies Including Greenhouse Gas Reduction" that USPTO launched in 2009 (USPTO, 2009).

[5] The Y02 class can be displayed and is searchable via http://worldwide.espacenet.com/classification?locale=en_EP#!/CPC=y02 (Veefkind *et al*., 2012, at p. 111, n12.)

[6] The total number of patents tagged with "Y02" in USPTO was 152,983 on October 25, 2013. The total number of patents tagged "Y02E 10/54$" was 5,021.

[7] Numbers retrieved on September 20, 2013. The column adds up to 5,764 based on 5,021 patents because of the possibility of double tagging of the same patents.

[8] Numbers retrieved on October 14, 2013; at this date the categories Y02E10/547 and Y02E10/549 were still incomplete in the PatStat database.

[9] At the date of development of the database (Aug. 20, 2013), we retrieved only 419 patents from this database.

We retrieved 419 patents at USPTO (on August 20, 2013) and 3,428 patents in PatStat (using the version of April 2013) with the CPC "Y02E10/541".[10] Figure 1 provides the trends.

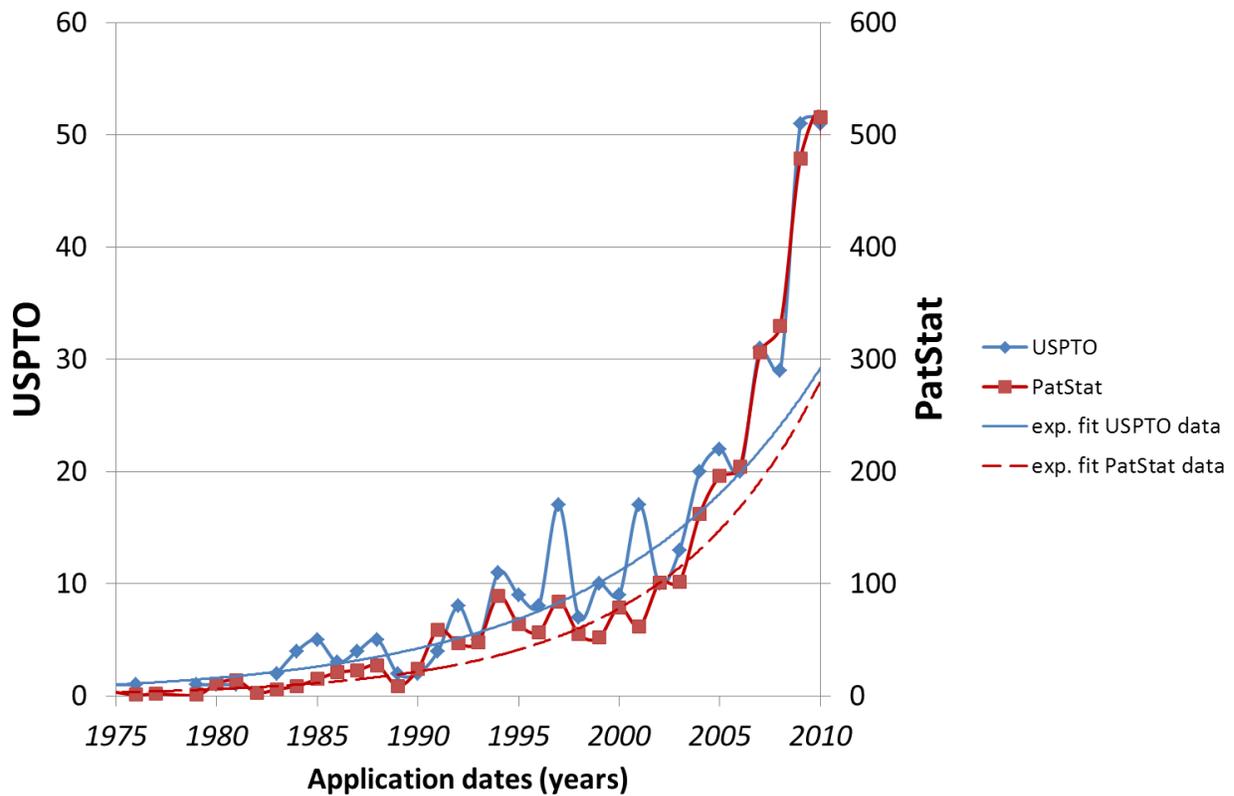

**Figure 1**: Development of patenting in USPTO and PatStat under the CPC tag Y02E10/541 for "CuInSe2 material PV cells", 1975-2010.

---

[10] Table 1 provides a number of 422 for the retrieval on 20 September 2013, but we use the 419 patents first downloaded from USPTO on 20 August 2013.

The attribution of this class in PatStat (right vertical axis) is an order of magnitude larger than in USPTO (on the left vertical axis). This difference accords with the expectation specified above: PatStat data contain duplicates from different patent offices. One can use priority patents to prevent this (de Rassenfosse *et al.*, 2013), but we use PatStat data in addition to USPTO data also for studying the geographical diffusion in markets other than the U.S. (Heimeriks *et al.*, in preparation). In other words, all patents at all offices are counted in the PatStat analysis, leading to double-counts, i.e. the actual number of different priority patents is smaller.

**4. Methods**

In this section, we discuss the routines and provide instruction on how to use the software that is freely available online for generating geographic maps (Section 4.1) and classification maps (Section 4.2).

Existing routines for overlaying patent data to Google Maps (Leydesdorff & Bornmann, 2012) and a map based on aggregated citations among IPC (Leydesdorff, Kushnir, & Rafols, 2012) were initially further developed for the purpose of *dynamic* mapping. The resulting routines are available at http://www.leydesdorff.net/software/patentmaps/dynamic for USPTO data and at http://www.leydesdorff.net/software/patstat for PatStat data. (These webpages also provide instructions about how to generate the various files.) The USPTO interface is accessed online by the routines, while the PatStat data have to be exported from a local installation of the database by using the dedicated scripts provided in SQL. The interface with USPTO additionally allows downloading the forward citations.

Unlike USPTO data, forward citation information in PatStat data is not uniformly standardized because references are provided by different patent offices. Considering citations from different offices raises questions about bias, as (at least part of) the citation could be due to differences in office practices and regulations, rather than to the quality and relevance of the patents considered (Criscuolo, 2006; Squicciarini *et al*., 2013, p. 8). Colors indicating citation counts above or below expected citation rates are therefore only provided when mapping USPTO data. As specified more extensively in Leydesdorff & Bornmann (2012), the proportion of top-cited patents in a sample of USPTO data can be ($z$-)tested for each location against the expectation, but only in the case of more than five patents at a city-location. As in the previous study, we test against the expectation that 25% of the patents at a location, *ceteris paribus*, can be expected to belong to the top-25% most-highly cited of the set.

Using colors similar to those of traffic lights, cities with (USPTO) patent portfolios significantly below expectation in terms of citedness are colored dark-red and cities with portfolios significantly above expectation dark-green. Lighter colors (lime-green and red-orange) are used for cities with an expected number of patents smaller than five (which should not statistically be tested) and for non-significant scores above or below expectation (light-green and orange).[11] (See at http://www.leydesdorff.net/photovoltaic/cuinse2/cuinse2_inventors.htm for the aggregated set.) The precise values are provided in the descriptors which can be accessed by clicking on the respective nodes. Additionally, all numerical values are stored in the file "geo.dbf" for statistical analysis.[12]

---

[11] This colour scheme was first used by Bornmann & Leydesdorff (2011) for *z*-testing proportions of publications in cities.

Data from PatStat are not *z*-tested in terms of citation rates, but rated in terms of percentiles of the patent distributions. Using a different color scheme (that is, the same colors as used by Bornmann *et al*. [2011]), the top-1% cities are in this case colored red (as "hot spots"), the top-5% fuchsia, the top-10% pink, the top-25% orange, the top-50% cyan, and the remainder (bottom-50%) is colored blue ("cold"). The percentile classes are relative to the specific years or sets of years under study.

*4.1    Geographic maps*

The user is first prompted to choose between an analysis of the address information of either inventors or assignees for the generation of geographic overlays. The addresses are then aggregated at the city level as provided in the patents. Using USPTO data, the addresses are almost always complete and standardized in the case of granted patents, but much less so in the case of patent applications. We use granted patents for this reason, but all time-series are organized in terms of the (earlier) filing dates.

PatStat data are drawn from different (e.g., national) databases and therefore heterogeneous in terms of the organization and quality of the address information. Our routines try to exhaust this data, but correction of error remains an uphill battle. Among the corrections to systematic error, we notably tried to correct for the state information when this is provided for addresses in the USA because the same city names may occur in different states (e.g., Athens, GA or Athens, OH). Several such minor adjustments are made automatically by the routine and we intend to improve this error-correction further.

---

[12] Differences between cities can also be *z*-tested for their significance as explained in Bornmann, De Moya-Anegón & Leydesdorff (2011). An Excel sheet available at http://www.leydesdorff.net/scimago11/index.htm can be used as guidance to this application.

In both cases (USPTO and PatStat), the addresses are first listed and have to be geocoded (for example, at http://www.gpsvisualizer.com/geocoder/ ).[13] Co-occurrence matrices of the addresses at the patent level are then generated for each year (or period of years). After completing this for the aggregated set(s), the new routines provide filters that allow the user to generate overlays to Google Maps for compilations of moving aggregates of years or single years. Because of the low numbers in the first decades (Figure 1), we used overlapping periods of five years in this study, as follows: 1974-1978; 1975-1979; 1976-1980; etc. However, the user can choose another time frame.[14]

The routines for both USPTO and PatStat data produce time-series of output files[15] that can be used as input for the generation of overlays to Google Maps at http://www.gpsvisualizer.com/map_input?form=data or a dedicated interface at http://data2semantics.github.io/PatViz. This latter site provides dynamic loading, visualization, and animation of the patent data using the JavaScript libraries of jQuery (http://jquery.com) and Google Maps (https://developers.google.com/maps/). This eliminates a number of steps in producing the visualizations. The resulting animations can be saved locally and made available at one's own website. The source code and program of PatViz are available for download at https://github.com/Data2Semantics/PatViz/releases; one can use this version locally and/or at the internet (see Appendix 1 for further instructions).

---

[13] The Bing Geocoder is also available from the Sci2 Tool at https://sci2.cns.iu.edu. This workflow is faster, but requires reformatting of the data (Sci2 Team, 2009). One can register for a free API key of Bing Maps at http://msdn.microsoft.com/en-us/library/ff428642.aspx .
[14] USPTO is available as html for patents granted since 1976; but the filing dates can be from earlier years. Our data begin in 1974 and the last period of five years is 2008-2012.
[15] The files are consecutively numbered as z1974.txt, z1975.txt, z1976.txt, etc. in the case of USPTO data—the "z" indicates that this data is *z*-tested—and pat1974.txt, pat1975.txt, pat1976.txt, etc., for PatStat data.

The routines also write a series of files (paj1974.txt, paj1975.txt, paj1976.txt, etc.) as input for network analysis using Pajek or any other network-analysis program reading the Pajek format.[16] These files contain symmetrical co-inventor (or co-assignee) data among cities in matrix format. One can use these files for generating network statistics such as density, degree distributions, etc., both for each year (or period of years) and over time.

## 4.2 Classification maps

For mapping the classifications, we use the base maps of aggregated citation relations among IPC in the USPTO data 1975-2011 provided by Leydesdorff, Kushnir, and Rafols (2012). These maps are available at http://www.leydesdorff.net/ipcmaps for both three and four digits of the current IPC version 8. We can use these maps for CPC because the first four digits of IPC were kept in the CPC scheme.

The initial step for the construction of the time-series is again the construction of the overall map for the aggregated set. Subsequently, the time series are generated by setting filters for consecutive years to this aggregate. In the case of USPTO data, the routine ipcyr.exe (available at

---

[16] Pajek is a program for the analysis and visualization of large networks that is available for free academic usage at http://pajek.imfm.si/doku.php?id=download .

http://www.leydesdorff.net/software/patentmaps/dynamic) generates input information for consecutive years in the format of VOSviewer for the mapping (http:// vosviewer.com). Two time series of files are generated as input for the mapping for three and four digits of IPC, respectively. Another routine (ps_ipcyr.exe at http://www.leydesdorff.net/patstat ) provides the same functionality for downloads from PatStat.

Both routines additionally write a file "rao.dbf" which contains Rao-Stirling diversity for both three and four-digit IPC-based maps for each consecutive year (or set of years). Rao-Stirling diversity is a measure that takes into account both the variety and the disparity in a patent portfolio under study across the IPC classes. The indicator is defined as follows (Rao, 1982; Stirling, 2007):

$$\Delta = \sum_{ij} p_i p_j d_{ij} \qquad (1)$$

where $d_{ij}$ is a disparity measure between two classes $i$ and $j$—the categories are in this case IPC classes at the respective level of specificity—and $p_i$ is the proportion of elements assigned to each class $i$. As the disparity measure, we use $(1 - cosine)$ since the cosine values of the citation relations among the aggregated IPC were used for constructing the base map of three and four digits. Jaffe (1986, at p. 986) proposed the cosine between the vectors of classifications as a measure of "technological proximity". Using the file "rao.dbf" in Excel, the development of the (Rao-Stirling) diversity over time can be plotted. Can the development of diversity perhaps be used as a measure of technological change? (e.g., Anderson & Tushman, 1990).

The IPC-based maps of VOSviewer for the different years can be animated (e.g., in PowerPoint) given the base maps of the aggregate of citation relations among IPC classes of patents between 1975 and 2011. The overlays show the evolution in specific samples against a stable background. An example of such an animation for the 419 USPTO patents in terms of IPC3 is provided at http://www.leydesdorff.net/photovoltaic/cuinse2/cuinse2.ppsx. One can animate the webpages of the geo-maps in PowerPoint similarly using the add-on "LiveWeb" at http://skp.mvps.org/liveweb.htm.

## 5. Results

### 5.1   USPTO

We first discuss the results of the analysis of using the 419 patents downloaded from USPTO with the search string "CPC/Y02E10/541", and turn thereafter to the larger set of 3,428 records downloaded with this CPC from PatStat for the comparison.

#### 5.1.1.  Geographical diffusion

After proper editing of the html (e.g., webpage titles and insertion of one's API code of Google Maps), one obtains a series of maps in which the node sizes are proportionate to the logarithm of the number of patents. [We used $\log(n+1)$ in order to prevent cities with single patents from disappearing because $\log(1) = 0$.] As noted, the node colors correspond to the quality of the patents in terms of their citedness (see Leydesdorff & Bornmann, 2012). One can click on each node to find statistical details. (This statistical data is also stored in the file "geo.dbf" that is generated and overwritten in each run.) The links span a network of co-inventor relations among the patents.

For example, Figure 2 provides the set of USPTO patents in this class (Y02E 10/541) for the five-year period 2000-2004. The numbers of patents are often too small for significance testing, but one can see at a glance that the US is dominant (green-colored nodes) in this set in terms of both numbers and quality. In addition to the US, Japan and Europe have developed their own networks. (One can zoom in on the map at http://www.leydesdorff.net/photovoltaic/cuinse2/index.html.) During this period, international co-inventorship between the three world regions was limited to transatlantic collaborations.

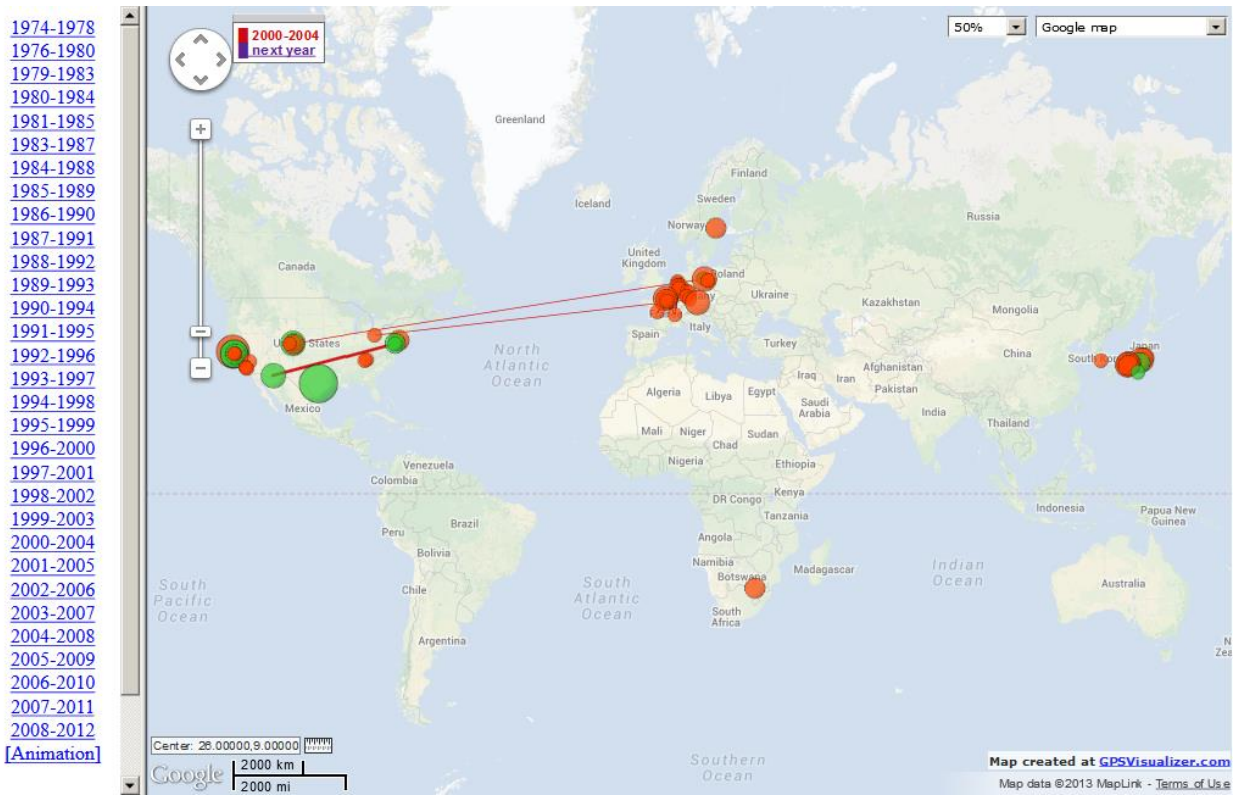

**Figure 2**: Patent configuration during 2000-2004 for CuInSe2 material in PV Cells (Y02E-10/541) in USPTO data; an interactive version of this map is available at http://www.leydesdorff.net/photovoltaic/cuinse2/index.html . See also at http://data2semantics.github.io/PatViz .

One can animate the map online by repeatedly clicking on the button "next year" to the right of the arrows of Google Map or by clicking on the button entitled "[Animation]" at the bottom left. (Alternatively, one can enter http://www.leydesdorff.net/photovoltaic/cuinse2/animate.html into the browser.) The animations require the reloading of the html—using a "refresh"—after each year and therefore run most reliably under a light browser such as Google Chrome.

As noted, we took a further step on the basis of this exploration and generated a dynamic interface for users at http://data2semantics.github.io/PatViz . In addition to showing the dynamics for this case study (and for its equivalent using PatStat data; see below), the interface allows users to upload their own geo-coded output files (z*.txt in the case of USPTO data or pat*.txt in the case of PatStat data) and to have generated the animations locally and/or at the Internet (Appendix 1).

Inspection of the animations informs us that patenting in this CPC class began in isolated centers in the USA, then spread first within the U.S. and thereafter also to some centers in Europe (e.g., 1983-1987). During the second half of the 1980s, Japanese and also isolated inventors in Europe began to patent in the USA. In 1990-1994, co-inventorship is found only in the local environments of Munich (Germany) and within Colorado. The latter network reflects that the National Renewable Energy Laborarory (NERL) of the US Department of Defense is based in Golden, Colorado. (NREL performs research on photovoltaics (PV) under the National Center for Photovoltaics.)

In the second half of the 1990s, there is also more co-invention in the USA and Japan, but within national boundaries. The technology increasingly becomes commercially viable during this period. The number of cities in Europe and Japan with USPTO patents increases, and transatlantic collaboration is resumed towards the end of the 1990s. Since 2003—the commercial phase—one sees co-invention between Japan and the USA, and within Europe. In the European context, France plays a role in addition to a recurrent collaboration between Germany and Spain. An address in the UK (Stirling in Scotland) joins the US networks in the final periods (2007-2011, 2008-2012). During 2008-2012, Europe is otherwise no longer represented in USPTO data.

In summary, collaborations within nations are more important than international collaborations, but the majority of the inventors do not collaborate beyond local environments. (The addresses on the patents can also be the home addresses of inventors.) How can the map in terms of IPC-classes add to our understanding of these geographical dynamics?

5.1.2. IPC classes

Figure 3 shows the IPC-based map (three digits) for the same set of patents as used in Figure 2 (2000-2004). The technology originated during the 1970s in the category of "basic electric elements" and remained there during the next 15 years, but has spread during the 1990s into other domains of technology such as "spraying and atomizing" and machine techniques for making thin films in photovoltaic cells. This diffusion increases further during the 2000s. (An animation is provided at http://www.leydesdorff.net/photovoltaic/cuinse2/cuinse2.ppsx.)

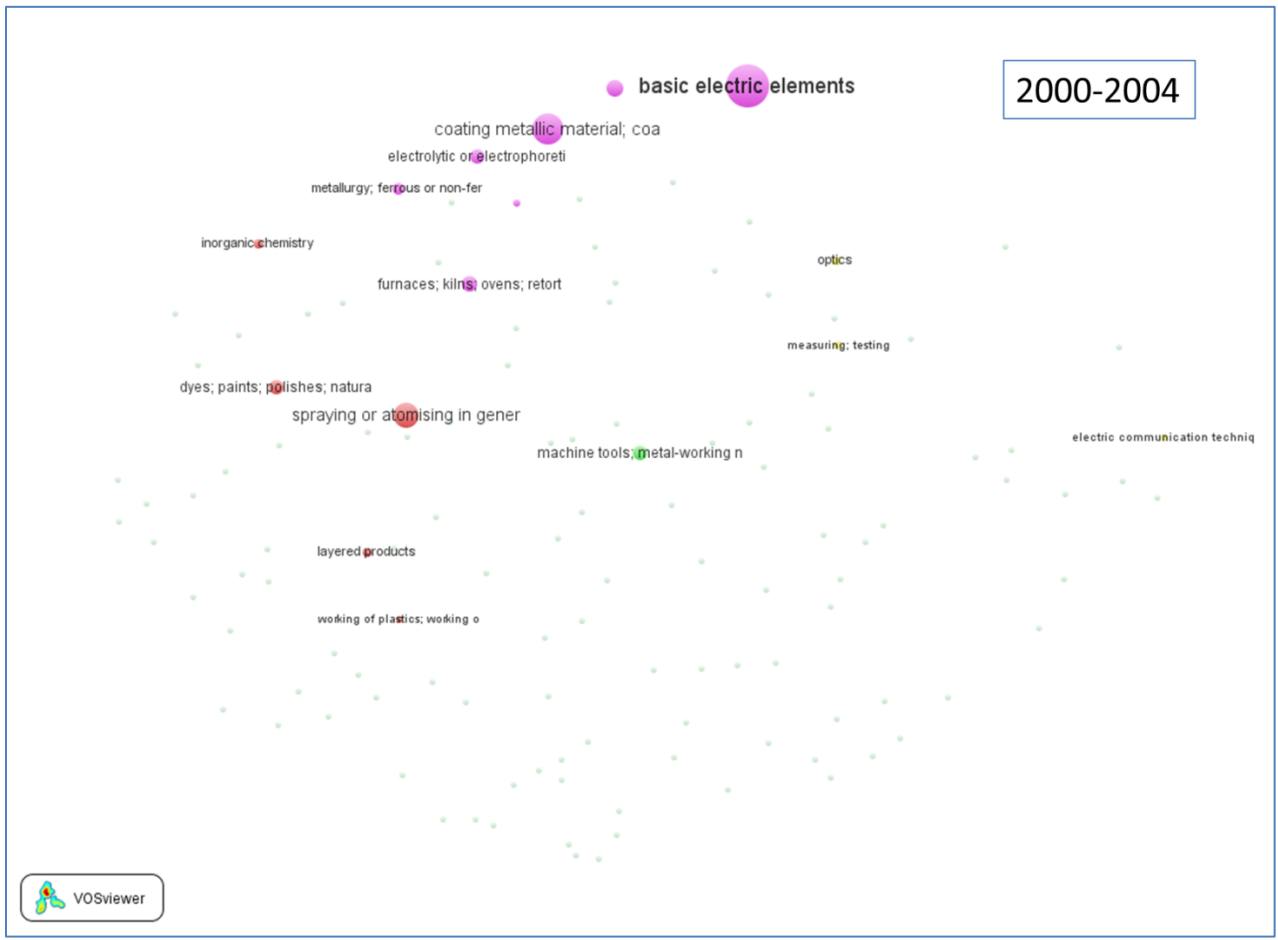

**Figure 3**: Map of USPTO patents in terms of IPC at the three-digit level for the period 2000-2004. A dynamic version of this map is available at http://www.leydesdorff.net/photovoltaic/cuinse2/cuinse2.ppsx.

Figures 2 and 3 can be combined into Figure 4 using frames in the html for the splitting of the screens (at http://www.leydesdorff.net/photovoltaic/cuinse2/dualmix.html). One can animate Figure 4 precisely as Figure 2. However, this animation taught us that dynamic changes in two different (split) screens are difficult to handle for an analyst. A user needs more control over the time steps when focusing on the *differences* between two dynamics. Therefore, we suggest another solution for studying the dynamics using split screens: by clicking on another year, one opens a new window in the browser with the same figures for this different year. A user is then able to compare among years using, for example, different time intervals (such as five or ten years) by going back and forth between windows, and at one's own pace.

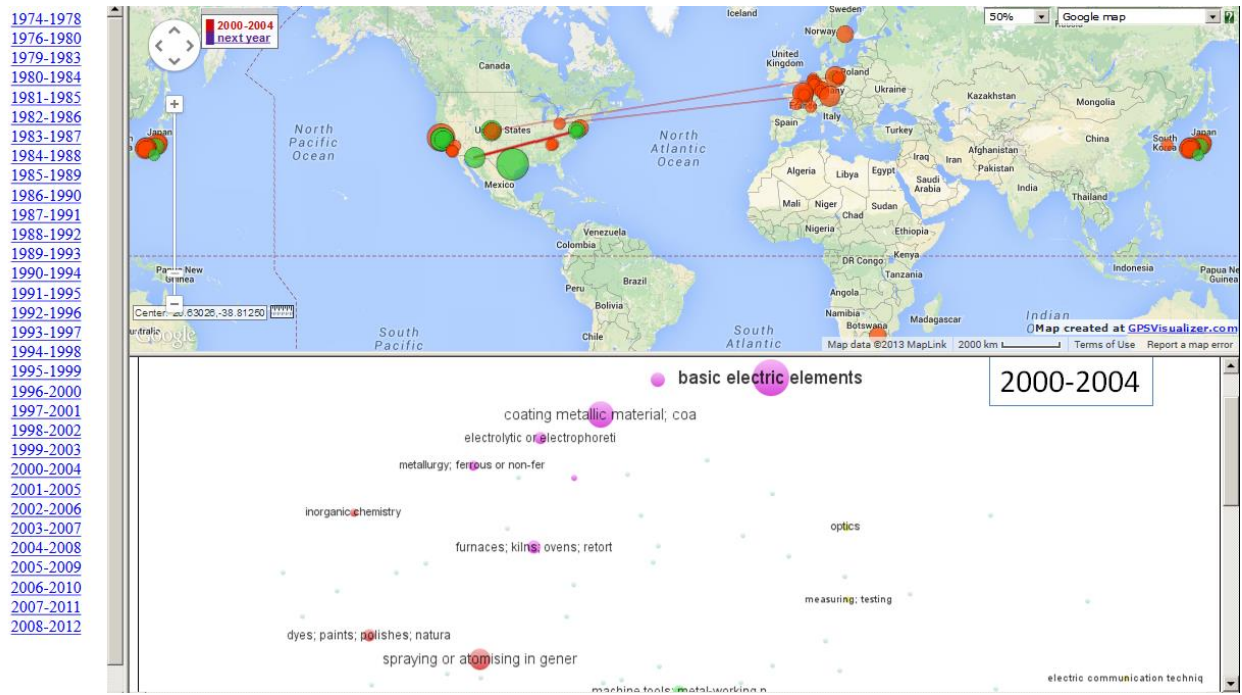

**Figure 4**: Map of USPTO patents in terms of both IPC (at the three-digit level) and geographical diffusion for the period 2000-2004; an interactive version of this map is available at http://www.leydesdorff.net/photovoltaic/cuinse2/dualmix.html .

Note that the maps in terms of the IPC classes (in the bottom half) can be enlarged to the full breadth of the screen by clicking on the map. We do not provide software for all possible combinations, but one can keep the html relatively simple so that a user can adapt the system to one's needs. The html of Figure 4, for example, reads as follows (Table 2):

```html
<html>
    <head>
        <title>USPTO (top) and IPC (bottom); inventors "CuInSe2 in material PV cells"; 1974-2012</title>
    </head>
  <frameset cols="*,8*">
  <frame src="frame1a.html">
  <frameset rows="50%,50%">
  <frame src="geo00.html">
  <frame scrolling="yes" src="v3_2000.png">
  </frameset>
</frameset>
</html>
```

**Table 2**: Html code for the two maps shown in Figure 4.

On larger screens, one would be able to show four or even more depictions in parallel. Thus, one would be able to study transitions which are visible in one domain in terms of other domains synchronically or also using different time frames. As noted in the introduction, the visualization of asynchronicities and development in different directions is central to our longer-term research program (Leydesdorff & Ahrweiler, in press; Leydesdorff *et al.*, 2013).

*5.1.3   Rao-Stirling diversity as a measure of technological change*

The longitudinal development of Rao-Stirling diversity indicates a cyclic pattern (Figure 5).

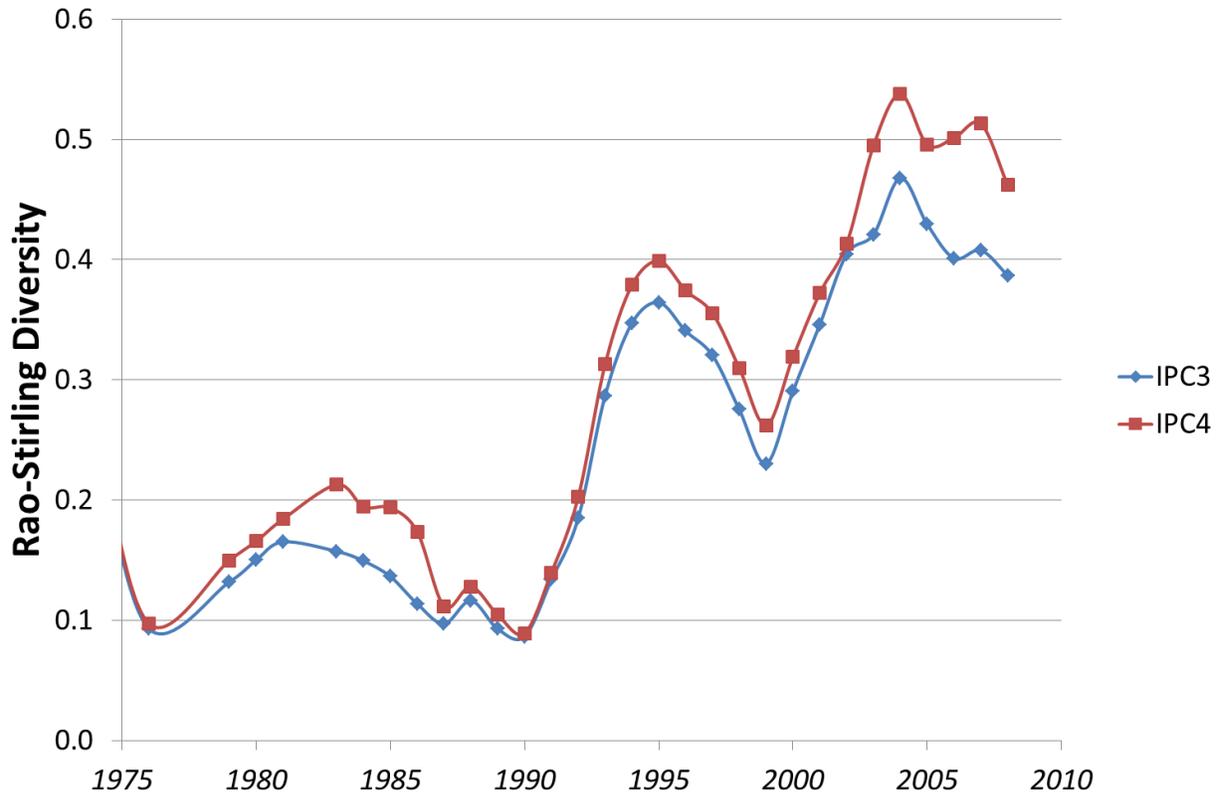

**Figure 5**: The development of Rao-Stirling diversity in IPC (three and four digits) among 419 USPTO-patents with CPC Y02E10/541 ("CuInSe$_2$ material PV cells") during the period 1975-2012.

Figure 5 suggests that the technology was developed in three cycles. Two of the valleys, i.e., the period of decreasing diversity in the late 1980s and the latest such period, correspond with breakthroughs in the efficiency of thin-film solar cells (Green *et al.*, 2013). Combining the maps with split-screens of Figure 4 for each consecutive year, we suggest specifying these cycles as follows (Shafarman & Stolt, 2003):

1. an early cycle during the 1980s which is almost exclusively American; after initial development of the technology at Bell Laboratories in the '70s, Boeing further developed the solar cells using these materials;

2. a second cycle during the 1990s that includes transatlantic collaboration and competition with Europe; the US, however, remains leading; and

3. a third and current cycle—the commercial phase—in which American-Japanese collaboration, on the one side, and collaboration *within* Europe, on the other, prevail.

The volume of patents continued to increase more smoothly (Figure 1), but with an increasing (above-exponential) rate during the most recent years. The pronounced articulation of these cycles in terms of Rao-Stirling diversity came as a surprise to us. As the material technology becomes mature, other technologies such as spraying the thin film on carrier materials may become crucial.

*5.2 PatStat*

We developed the same routines analogously for the patent data downloaded from PatStat. As noted, this data is an order of magnitude larger than in USPTO (Figure 1), since PatStat collects patent data from offices in different countries and world regions. The geographical map for the same year as used above (2000-2004) is provided in Figure 6. This figure can be animated similarly as in the case of USPTO data in Figure 2 above—that is, by clicking on the button entitled "[Animate]". This animation is also implemented in the JavaScript-based program PatViz at http://data2semantics.github.io/PatViz or http://www.leydesdorff.net/patviz (see Appendix 1).[17]

---

[17] The program itself and the source code can be downloaded at https://github.com/Data2Semantics/PatViz/releases.

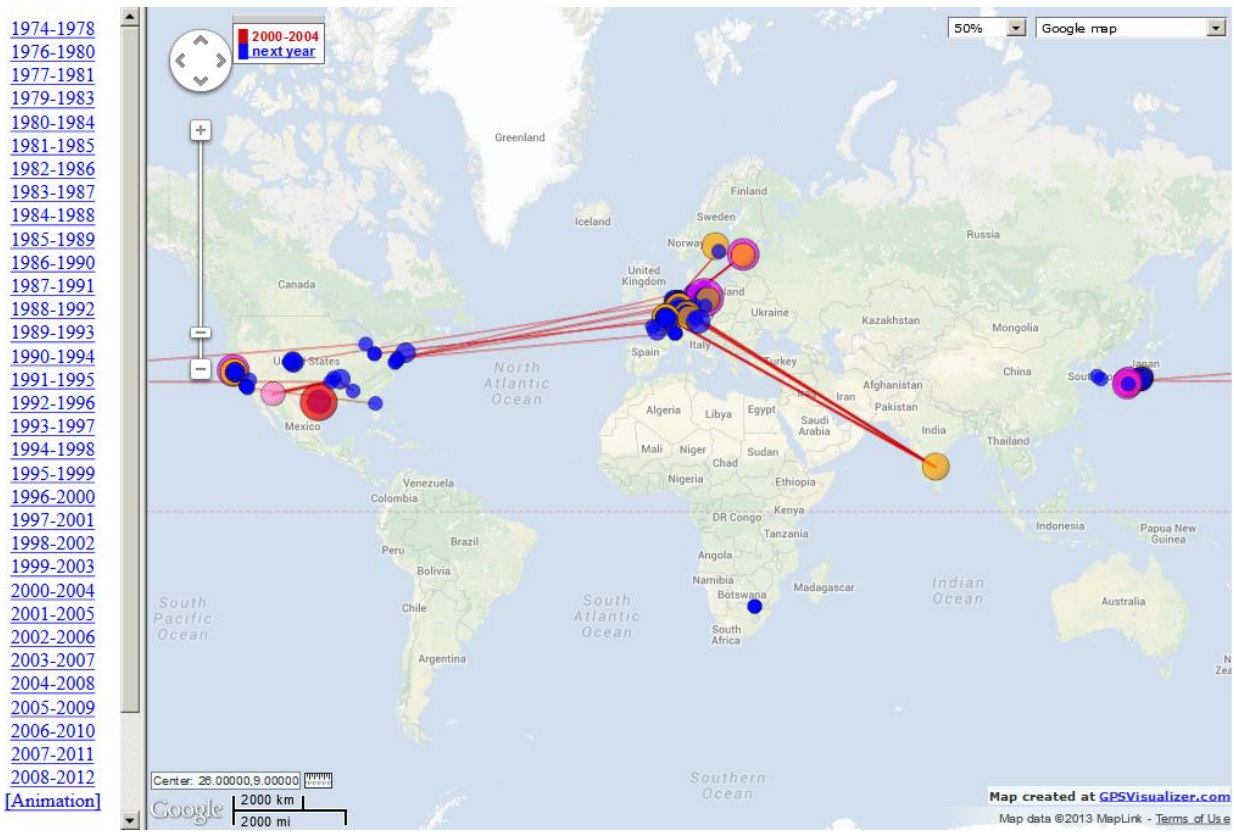

**Figure 6**: Patent configuration during 2000-2004 for "CuInSe$_2$ material in PV Cells" (Y02E-10/541) in PatStat data; an interactive version of this map is available at http://www.leydesdorff.net/photovoltaic/cuinse2.patstat/index.htm .

The colors in Figure 6 use a palette different from Figure 2 because this data cannot be assessed in terms of citations. In this figure, "red" means hot, and "blue" cold in terms of relative numbers of patents at locations (Bornmann *et al.*, 2011). Otherwise, the map is not very different from the one based on USPTO data (in Figure 2). The PatStat network can also be considered as an extension of the USPTO network. For example, the Indian center in Chennai is added. This center is well connected to leading centers in Germany and France.

In order to enhance the possibility to make comparisons, we experimented with a split screen showing the USPTO data in the top screen and PatStat data for the same year(s) at the bottom

(Figure 7; available online at

http://www.leydesdorff.net/photovoltaic/cuinse2.patstat/dualgeo.html). For the same reasons as above, we abstain from animating this double map because of overloading one's mental map, but instead the option is provided to compare for different years in terms of new windows in a browser.

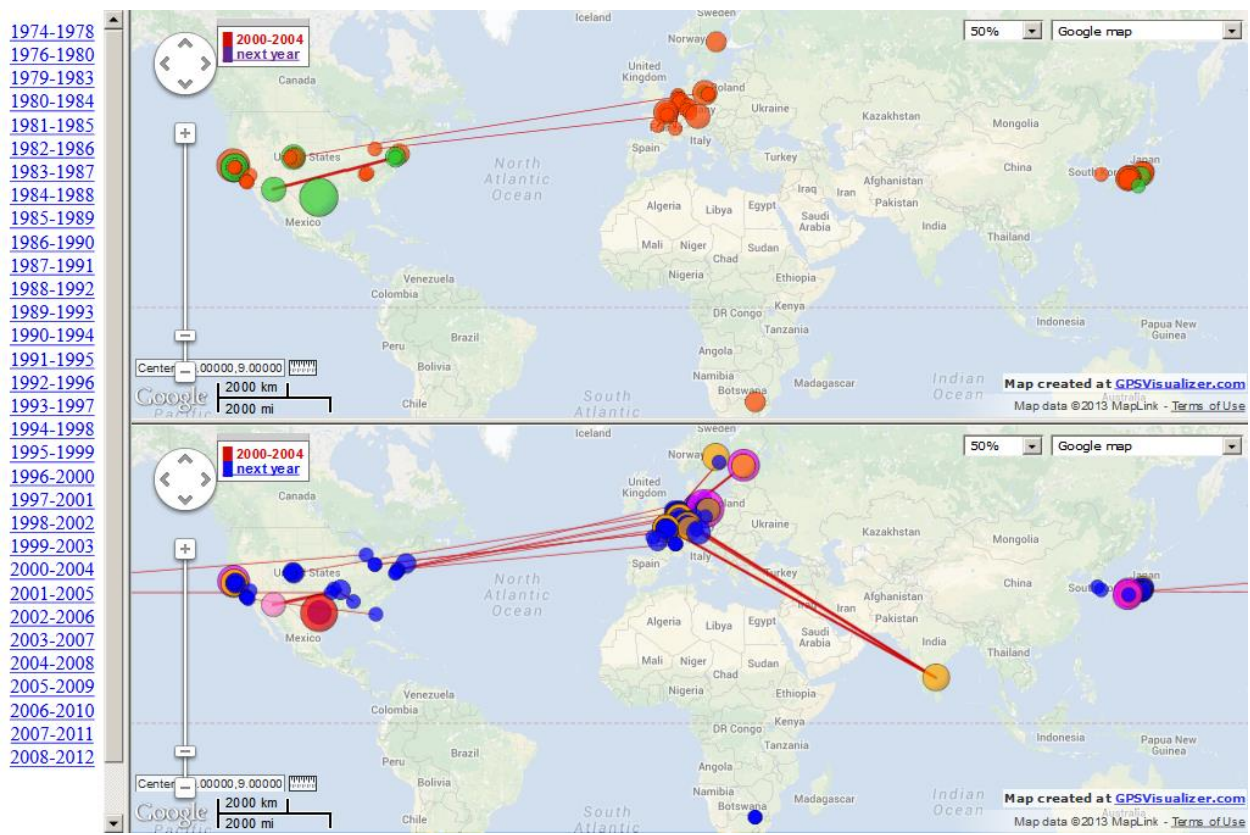

**Figure 7**: Comparison of USPTO-based and PatStat-based global maps of patents classified as "CuInSe$_2$ material PV cells" (CPC); an interactive version of this map is available at
http://www.leydesdorff.net/photovoltaic/cuinse2.patstat/dualgeo.html

The juxtaposition of the geographical maps for USPTO and PatStat data for each year and over the years in separate windows enables an analyst to zoom into the differences and similarities. One can follow up with network analysis using the files in the Pajek format that are generated additionally by our routines. Figure 8 shows the largest network components during 2000-2004 in the two sets of patents classified with "CuInSe2 material in PV Cells" (Y02E10/541) and using the same data as in Figures 2 and 6 above. In addition to spelling variants and misspellings in the PatStat database such as Rueil-Malmaison (France)—with or without hyphen—and Jülich (Germany)—with or without umlaut—the two graphs show the extension of the network in PatStat including non-US patents.

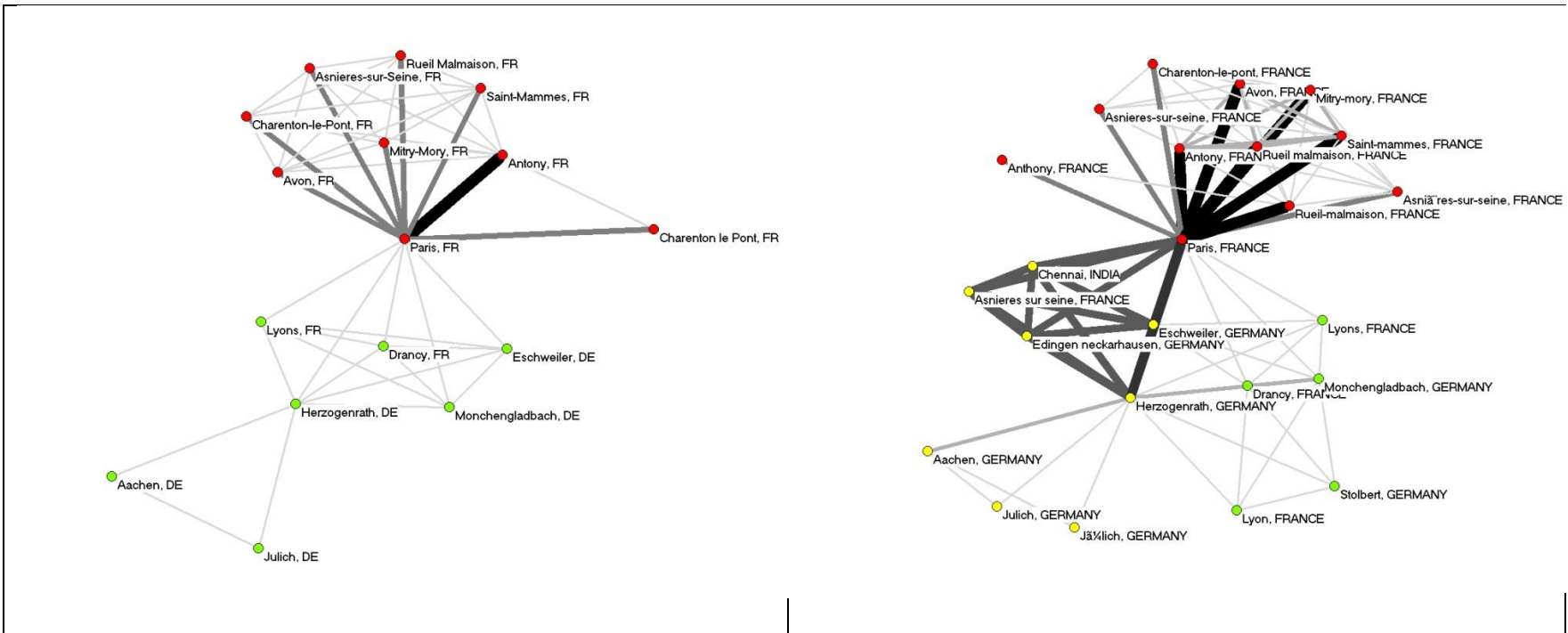

**Figure 8**: Largest components of the co-inventor network in USPTO and PatStat data for patents in "CuInSe2" (Y02E10-541) during 2000-2004; 16 and 24 nodes, respectively. Coloring of the community structure is based on the algorithm of Blondel *et al.* (2008); Kamada & Kawai (1989) is used for the layout.

In addition to this network of co-inventors between France and Germany, Figure 9 shows other (separate) networks in this same year among German, Dutch, and Japanese inventors, and one network with German, Dutch, Belgian, and Estonian participants (in the upper left-side corner). Note that US inventors are not networked internationally during this period (2000-2004).

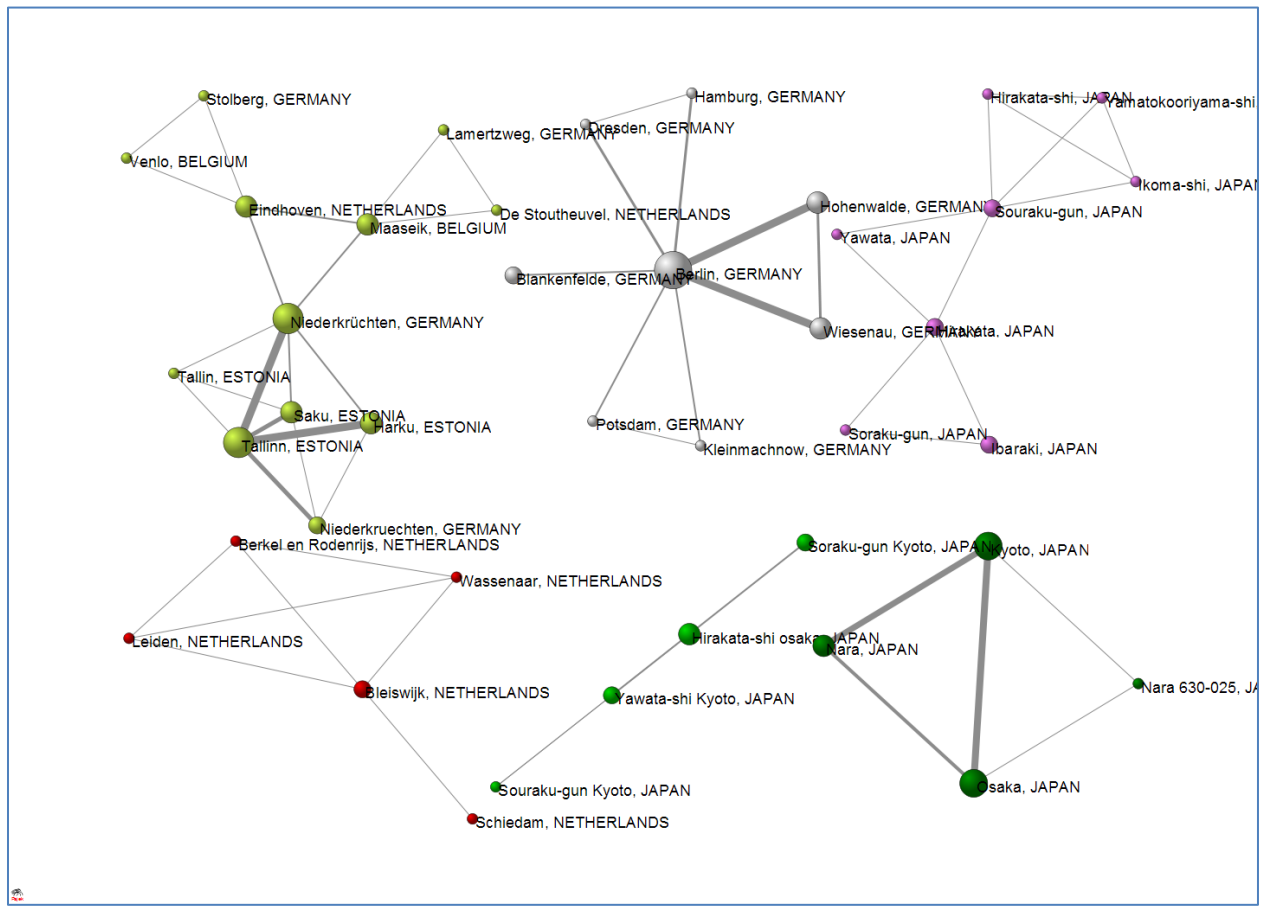

**Figure 9**: Components other than the largest one (see Figure 8) in the co-inventor network of patents in PatStat during 2000-2004.

Indeed, one would find a poor representation of these national and regional networks using USPTO data (Figure 2). When comparing the two overall networks for 2000-2004 in terms of various network parameters (e.g., De Nooy *et al*., 2005; Hanneman & Riddle, 2005), the density value is significantly different: density in the USPTO data 2000-2004 is twice as high as for PatStat data in this same period. Thus, while the PatStat network is larger in size, it is less densely connected than the USPTO network among inventors. The USPTO network can be considered as a core set within the larger network of PatStat data. The number of communities in this PatStat data is 67 as against 32 in USPTO data. Although this seems to support the idea of showing niche markets (e.g., in India), 47 of these groups are isolates, and thus most likely local duplicates.

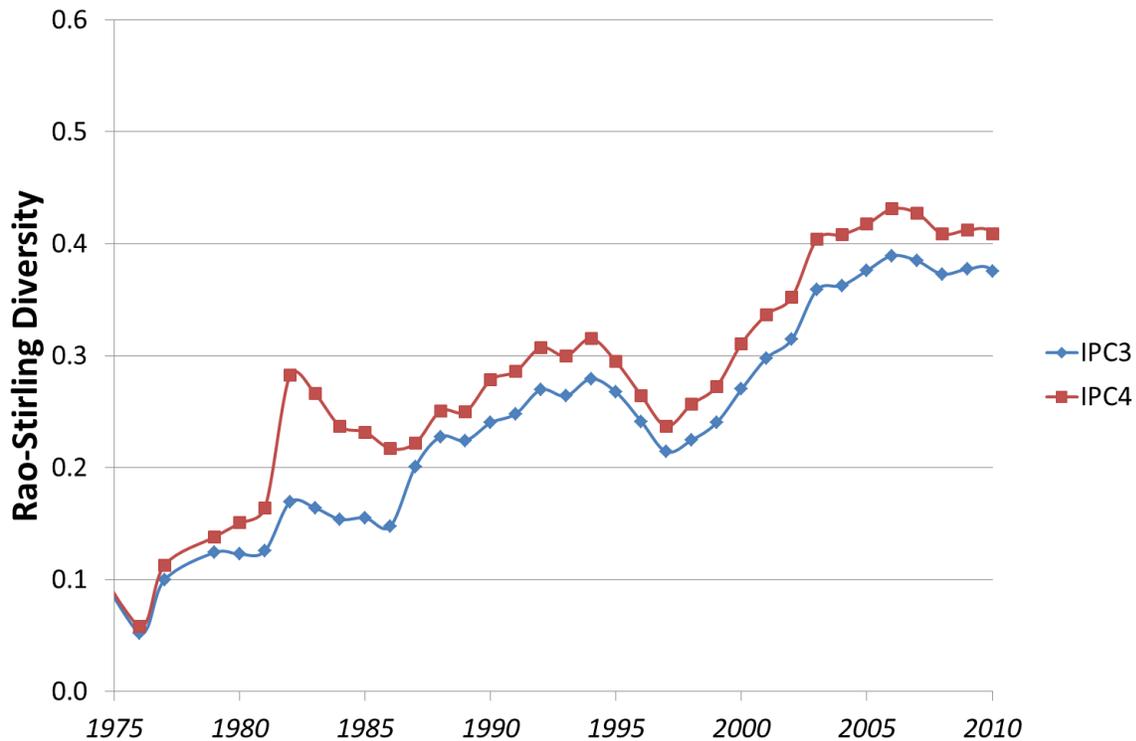

**Figure 10:** The development of Rao-Stirling diversity in IPC (three and four digits) among 3,428 patents in PatStat during the period 1975-2010.

Figure 10 shows the longitudinal development of Rao-Stirling diversity in the set of 3,428 patents downloaded from PatStat using the CPC of Y02E10/541. Note that Rao-Stirling diversity might be used as a rough first indicator of a possible "technological change," but not as an actual measure of this complex phenomenon. However, one can distinguish the same three cycles of development as in Figure 5, but less pronounced when compared with USPTO data. This accords with the expectation because PatStat includes national databases which may experience the various cycles with more delays than among patents in USPTO. The shift to a next generation of the technology is provoked by sharp competition in the US market, but not necessarily followed in more protected market environments in other nations or world regions. In other words, one can expect the diffusion patterns to develop more gradually using PatStat data because of this effect of averaging out among the different sources of patent data. New generations of patents may be delayed in the worldwide database of PatStat when compared with the more competitive environment of USPTO.

**6. Discussion about the longitudinal development of diversity**

Let us further explore our conjecture about technological generations made visible by time-series of Rao-Stirling diversity, by using the next following CPC category, that is, the class Y05E10/542 for "dye-sensitized solar cells" (DSSC). In Figure 11, Rao-Stirling diversity is plotted at the four-digit level for both USPTO and PatStat data. The data suggest at least two cycles: a first one that ran out of steam during the 1980s, and a second one during the 1990s. Perhaps, a third one can be distinguished as emerging in USPTO data during the most recent period, that is, since 2004.

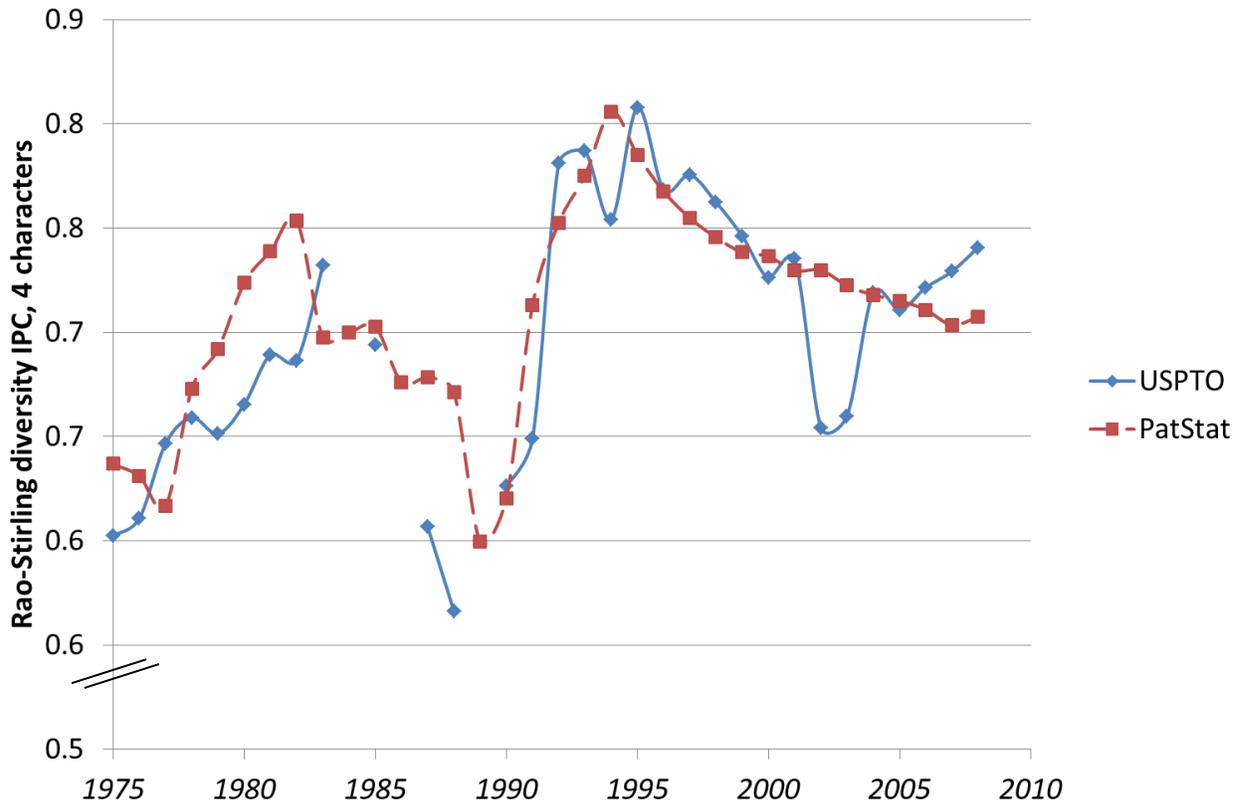

**Figure 11:** Development of Rao-Stirling diversity in Y02E10/542 ("Dye sensitized solar cells") for USPTO and PatStat, respectively.

The second wave (in the early 1990s) corresponds to the invention of the modern (second generation) version of DSSC which was developed in the period 1988-1991. The first highly efficient DSSC—also known as the Grätzel cell—was published in 1991 (O'Regan & Grätzel, 1991). A patent was filed at the World Intellectual Patent Organization according to the Patent Cooperation Treaty (PCT) in March 1993 (WO93/18532), and then also at USPTO in November 1993 (nr. 5,525,440 in USPTO; granted June 11, 1996). The École Polytechnique Fédérale de Lausanne is the assignee of this patent. Patenting, however, seems to have become broader in scope already a few years earlier (Schmookler, 1962); and shortly after 1993, the diversity begins to decline. The plots in Figures 5 and 11 provide us with heuristics for the reconstruction of the history of a technology from this data. In other words, informed questions can be raised and discussed with expert knowledge in the various domains.

Figures 5 and 11 may seem somewhat similar upon visual inspection (the Spearman rank correlations are .81 in the case of IPC3 and .48 in the case of IPC4), but in other cases we found significantly negative correlations, such as $\rho = -.71$ (p<0.01) between "microcrystalline silicon PV cells" (Y02E10/545) and "polycrystalline silicon PV cells" (Y02E10/546). An expert in PV research whom we consulted confirmed that these are very different technologies (Van Sark, *personal communication*, 7 January 2014). "Microcrystalline silicon PV cells" (Y02E10/545) are more similar to "amorphous silicon PV cells" (Y02E10/548) and "polycrystalline" (Y02E10/546) is more similar to "monocrystalline" (Y02E10/547). The Spearman rank correlation between these last two time-series, for example, is 0.69 ($p < 0.01$).

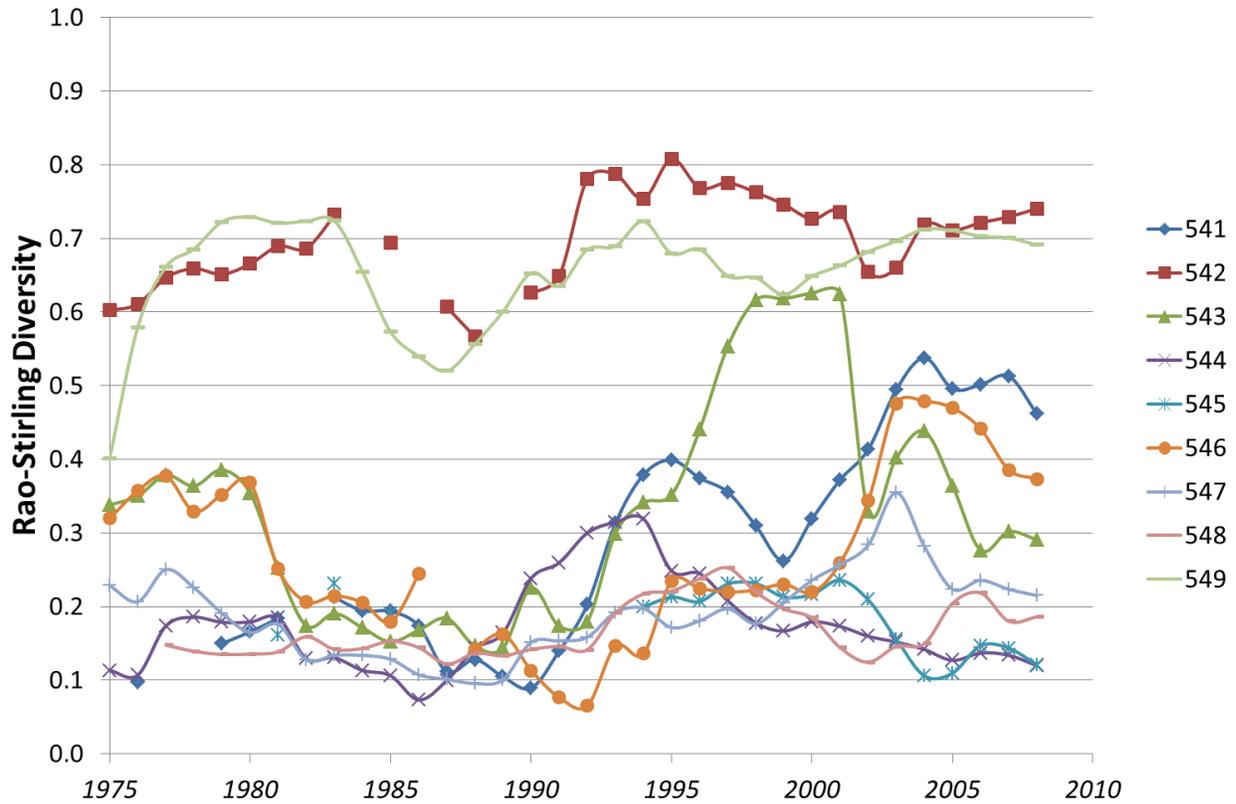

**Figure 12**: The development of Rao-Stirling diversity in USPTO-patents of nine material technologies for photovoltaic cells.

Figure 12 shows the results for an extension to the nine material technologies classified as Y02E10/54*. These results merit further investigation and interviews for validation with experts in the respective fields.

## 7. Conclusions

The maps of patents in different dimensions are instrumental to understanding the complex dynamics of innovation by providing different projections of these dynamics. We distinguished in this study between IPC-based maps that show the technological organization of the patents in a vector space, the geographic maps as overlays to Google Maps, and the social networks that can be overlaid to the geographic map, but can also be studied in themselves using graph-theoretical instruments such as spring-embedded layouts (e.g., Kamada & Kawai, 1989; see Figure 8 above).

The user, or more generally the discourse of innovation studies, can bring the insights that can be harvested from the different perspectives together reflexively. The maps provide the footprints of the development; but they can make the historical narrative evidence-based. We elaborated this for the case of $CuInSe_2$ as a material technology for photovoltaic cells. At the theoretical level, we thus aim to address what Griliches (1994) called "the computer paradox," but from a methodological angle: ever more data—nowadays, one would say "big data"—are stored in ever larger repositories. The logic of these repositories is institutional, whereas the logic of innovation is based on the transversal recombination of functions at interfaces (e.g., supply and demand). The relabeling using the Y-tag in CPC, however, provides an opportunity to follow delineated technologies within and across databases: recent agreements of EPO and USPTO with the Chinese, Korean, and Russian patent offices to use also CPC in the near future show an increased awareness to coordinate the data in a networked mode.

The advantage of developing instruments is provided by the direct relation between instruments such as visualization and the empirical operationalization (McGrath *et al.*, 2003). Middle-range theorizing can guide this process of developing "instrumentalities" (Price, 1984) as heuristics (Geels, 2007). The systems perspective adds the evolution of these functions over time in terms of technological trajectories and regimes (Arthur, 2009). Empirical studies of innovation need to allow for the appreciation of changes of perspectives because innovations can be developed—or unintentionally diffuse—into different directions: geographical, economic, and technical. In our opinion, the bottle neck of innovation studies has been the development of instruments which keep pace with the (re)combinations possible in terms of the data fluxes.

Dynamic overlays that can be accessed interactively on the internet provide the user with options to trace technological developments and develop new perspectives reflexively. The use of Rao-Stirling diversity in this study can be considered as a case in point: the literature pointed us to considering variety versus the loss of variety in shake-out phases as central to techno-economic developments (Anderson & Tushman, 1990; He & Fallah, 2011), but the data allowed us to operationalize this in relation to the new instruments. The extension beyond two maps to be recombined follows as a progressive research agenda for quantitative innovation studies (Rotolo *et al*., in preparation).

**Acknowledgements**
We are grateful to comments on a previous version of this manuscript by Jan Youtie (Georgia Tech) and Wilfried van Sark (Utrecht University). We also thank two anonymous referees for their constructive comments.

**Appendix I**

The PatViz tool enables users to animate output from the (geo-coded) patent maps produced from USPTO data (at http://www.leydesdorff.net/software/patentmaps/dynamic) or from PatStat data (at http://www.leydesdorff.net/software/patstat) both locally and online. Interactive versions are provided at http://www.leydesdorff.net/patviz or http://data2semantics.github.io/PatViz.

Instead of generating and visualizing the maps one by one for each year consecutively (at http://www.gpsvisualizer.com/map_input?form=data), PatViz reads an entire time series of files first. Using JavaScript, the program automatically generates the animation using the same parameters as specified in this study. Currently, one can be upload files with the names pat*.txt (e.g., pat1980.txt, pat1981.txt, etc.) as generated by ps_geoyr.exe for PatStat data; and the files z*.txt generated by usptoyr.exe for USPTO data. Instructions for preparing these files can be found at http://www.leydesdorff.net/software/patstat and http://www.leydesdorff.net/software/patentmaps/dynamic, respectively.

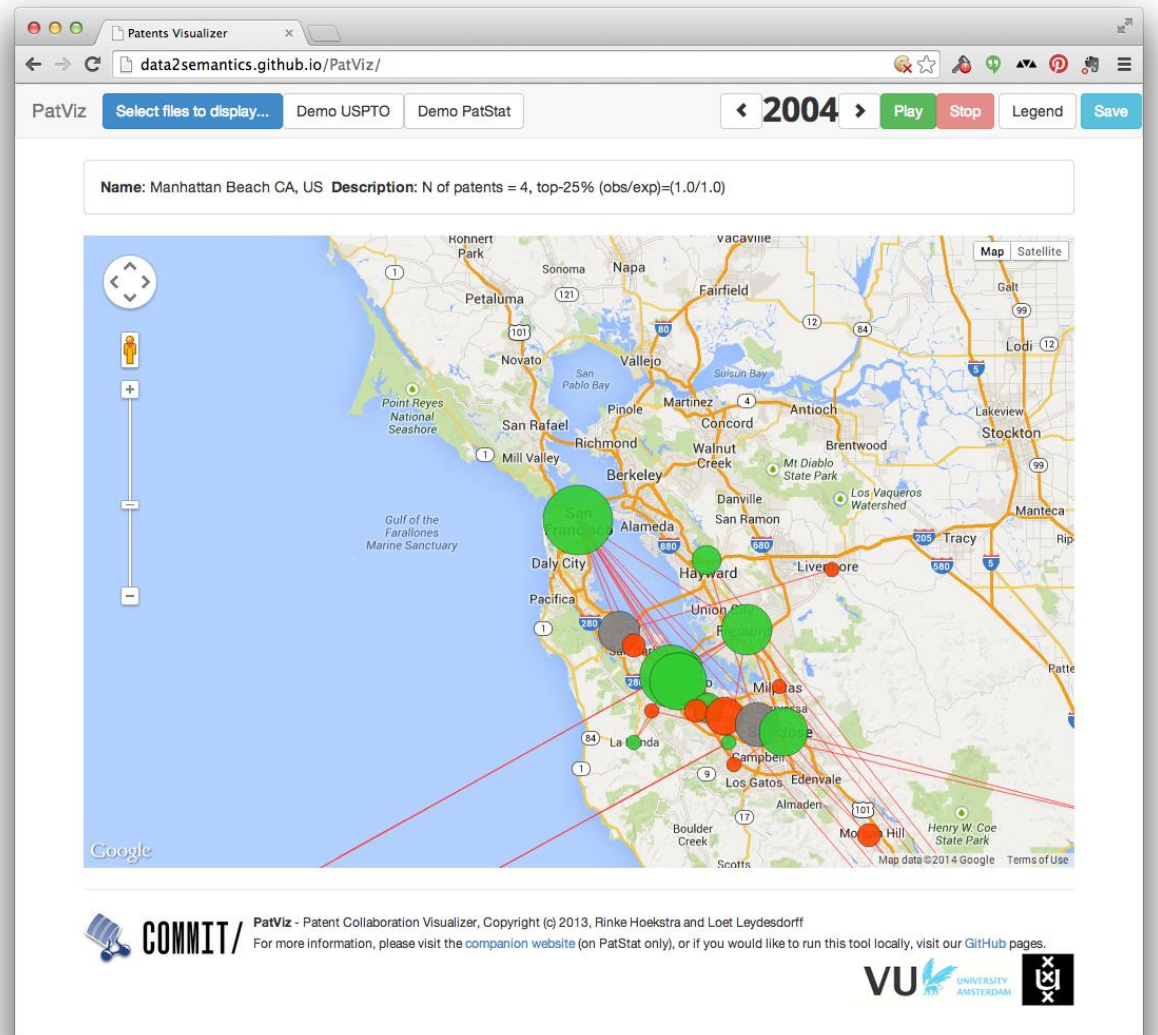

**Figure A1:** Screenshot of the PatViz interface, focusing on the Bay Area.

Users can load their own data files by clicking "Select files to display...", the two demo buttons provide access to data for $CuInSe_2$ as material technology for PV cells retrieved on the basis of Y02E10/541 as the Cooperative Patent Classification for the download in USPTO and PatStat, respectively. Figures 2 and 5 above provide snapshots of these two configurations in 2000-2004. For an example and further instructions, see http://www.leydesdorff.net/photovoltaic/patviz.

Users can scroll through the years by clicking the "<" and ">" buttons in the menu bar. Clicking "Play" will start an animation that will automatically cycle through all years available in the dataset. Clicking "Stop" halts the animation. The "Legend" button gives information about the colors used in the visualization.

At the top right of the screen, the "Save" button enables users to save the results for demonstration purposes in a single html-file (containing the specific data set) that can be run locally using a browser, or hosted online. After clicking "Save", PatViz prompts for a filename and for a Google Maps API key (that is freely available from Google at [http://console.developers.google.com](http://console.developers.google.com)) so that all interfaces are available; an Internet connection remains required for this application since it depends on externally hosted JavaScript libraries.

The latest release of PatViz can be downloaded from [https://github.com/Data2Semantics/PatViz/releases](https://github.com/Data2Semantics/PatViz/releases) for installation at one's own machine. After unzipping the files, one installs the program and can run it by opening the index.html file in a Web browser. The program requires that the computer be connected to the Internet in order to download the Google Maps and other external libraries. The program can also be uploaded and used online, after replacing the API key of Google Maps in index.html with the one for one's own website.